\documentclass[lettersize,journal]{IEEEtran}
\usepackage{amsmath,amsfonts}
\usepackage{algorithmic}
\usepackage{algorithm}
\usepackage{array}
\usepackage[caption=false,font=normalsize,labelfont=sf,textfont=sf]{subfig}
\usepackage{textcomp}
\usepackage{stfloats}
\usepackage{url}
\usepackage{enumitem}
\usepackage{verbatim}
\usepackage{graphicx}
\usepackage{cite}
\usepackage{array}
\usepackage{color}
\usepackage{graphicx}
\usepackage{caption}
\usepackage{float}
\usepackage{cleveref}
\Crefname{figure}{Fig.}{Figs.}

\begin{document}
\title{Shadow Wireless Intelligence: Large Language Model-Driven Reasoning in Covert Communications}

\author{Yuanai~Xie, Zhaozhi~Liu, Xiao~Zhang, Shihua~Zhang, Rui~Hou, Minrui~Xu*, Ruichen~Zhang, \\ and Dusit~Niyato,~\IEEEmembership{Fellow,~IEEE}

\thanks{\emph{(Corresponding author: Minrui Xu.)}}
\thanks{Y.~Xie, Z.~Liu, X.~Zhang, S.~Zhang, and R.~Hou are with the School of Computer Science, South-Central Minzu University, Wuhan 430074, China. Emails: 2023002@scuec.edu.cn, 2024120415@mail.scuec.edu.cn, xiao.zhang@my.cityu.edu.hk, zhangshihua@scuec.edu.cn, hourui@scuec.edu.cn.}
\thanks{M.~Xu, R.~Zhang, and D.~Niyato are with the College of Computing and Data Science, Nanyang Technological University, Singapore. Emails: minrui001@e.ntu.edu.sg, ruichen.zhang@ntu.edu.sg, dniyato@ntu.edu.sg.}
}

% The paper headers
%\markboth{}

\IEEEpubid{}
% Remember, if you use this you must call \IEEEpubidadjcol in the second
% column for its text to clear the IEEEpubid mark.

\maketitle
\begin{abstract}
Covert Communications (CC) can secure sensitive transmissions in industrial, military, and mission-critical applications within 6G wireless networks. However, traditional optimization methods based on Artificial Noise (AN), power control, and channel manipulation might not adapt to dynamic and adversarial environments due to the high dimensionality, nonlinearity, and stringent real-time covertness requirements. To bridge this gap, we introduce Shadow Wireless Intelligence (SWI), which integrates the reasoning capabilities of Large Language Models (LLMs) with retrieval-augmented generation to enable intelligent decision-making in covert wireless systems. Specifically, we utilize DeepSeek-R1, a mixture-of-experts-based LLM with RL-enhanced reasoning, combined with real-time retrieval of domain-specific knowledge to improve context accuracy and mitigate hallucinations. Our approach develops a structured CC knowledge base, supports context-aware retrieval, and performs semantic optimization, allowing LLMs to generate and adapt CC strategies in real time. In a case study on optimizing AN power in a full-duplex CC scenario, DeepSeek-R1 achieves 85\% symbolic derivation accuracy and 94\% correctness in the generation of simulation code, outperforming baseline models. These results validate SWI as a robust, interpretable, and adaptive foundation for LLM-driven intelligent covert wireless systems in 6G networks.
\end{abstract}


\section{Introduction}
The advent of 6G wireless communication technologies has drastically increased the openness and complexity of wireless transmissions, posing new security challenges to sensitive information transmissions. The advancement of AI in 6G, especially the emergence of Large Language Models (LLMs) due to massive data, enhanced computing performance, and the Transformer architecture \cite{Transformer}, has been significant \cite{EVOLUTION_OF_LLM}. In some sensitive scenarios, such as military, industrial, and next-generation IoT deployments, ensuring wireless transmissions remain undetectable has become crucial to counter advanced adversarial surveillance and interception attempts \cite{threat}. As a promising solution, Covert Communications (CC) typically rely on strategies such as Artificial Noise (AN) insertion, precise power control, and dynamic channel manipulation to reduce the warden's detection probability beneath rigorously predefined security thresholds \cite{Covert_Communication_comprehensive_survey}. By using CC techniques, such as low power and sparse signaling, CC systems can selectively engage only the most critical expert paths, optimizing energy efficiency and reducing computational overhead even under low Signal-to-Noise Ratio (SNR) conditions. Furthermore, by integrating adversarial-aware strategies like dynamic channel state estimation and counter-response prediction, CC systems can robustly counteract noisy inputs and hostile interference, ensuring reliable performance in operation.

However, traditional optimization approaches for CC exhibit limited adaptability in dynamic and adversarial environments, as they highly depend on explicit mathematical modeling and iterative solvers~\cite{covert_challenge}, which are often computationally prohibitive for real-time deployment in 6G networks. In contrast, LLMs~\cite{IAI_IN_NET} offer a more flexible alternative by interpreting task objectives and generating optimization strategies directly from natural language or structured input. However, standalone LLMs still face key limitations, including insufficient mathematical foundation, symbolic hallucinations, and outdated domain knowledge. To bridge this gap, CC can be leveraged not only as a target application but also as a rigorous testbed for advancing LLMs inference capabilities. Specifically, reinforcement learning-enhanced Mixture-of-Experts (MoE) \cite{MoE} LLMs, e.g., DeepSeek-R1, can be integrated into CC as a reasoning engine to enable real-time symbolic derivation, code generation, and context-aware optimization. \textbf{Sparse inference aligned with low-power CC settings is supported by its MoE-based routing, while its structured Chain-of-Thought (CoT) and Retrieval-Augmented Generation (RAG) enhance interpretability and semantic accuracy under adversarial constraints}. In this way, there is a bi-directional synergy where CC can improve LLMs robustness and reasoning precision, while LLM transforms CC into an adaptive, interpretable, and intelligent process—laying the foundation for secure and efficient wireless intelligence in 6G.

\begin{figure*}[t]
        \centering
\includegraphics[width=0.9\linewidth]{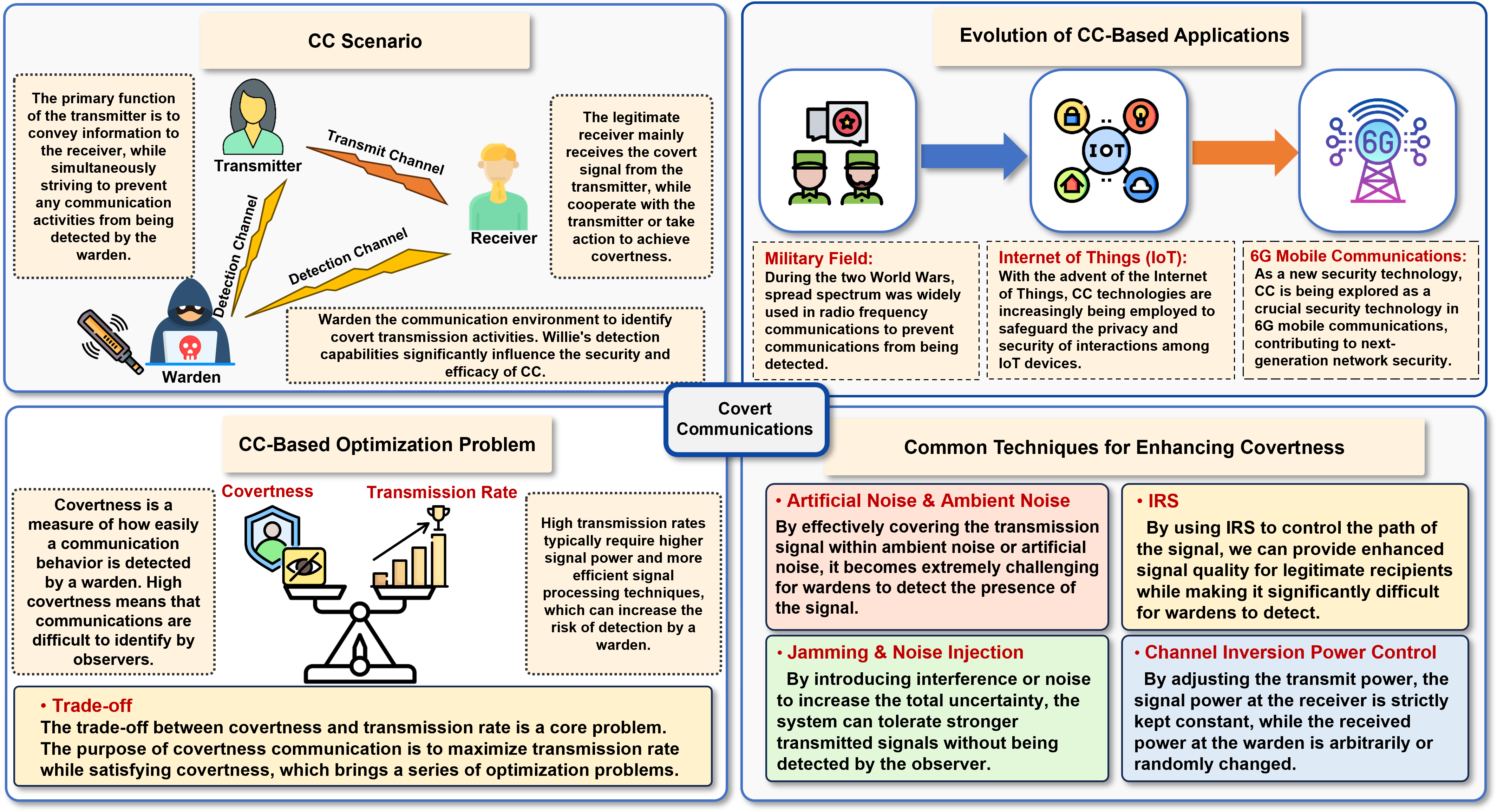}
        \caption{Overview of CC technologies: scenarios, evolution, optimization
        challenges, and enhancement strategies.}
        \label{fig:covet_communication_overview}
    \end{figure*}
    
    \indent
     To address these limitations, this article proposes the concept of Shadow Wireless Intelligence (SWI), integrating advanced LLM-driven reasoning with RAG \cite{RAG}, a method allowing real-time retrieval of specialized knowledge during inference. Specifically, we leverage DeepSeek-R1~\cite{guo2025deepseek}, a powerful MoE-based LLM that dynamically activates expert sub-networks tailored to specific input contexts, significantly improving symbolic interpretation and mathematical reasoning accuracy. In our proposed framework, DeepSeek-R1, enhanced with RAG, dynamically retrieves relevant mathematical formulas, theoretical specifications, and empirical results, thereby significantly augmenting contextual understanding and reasoning precision for CC-based optimization. By integrating advanced reasoning capabilities with dynamic knowledge retrieval, the SWI paradigm provides a new foundation for future intelligent, autonomous, and secure wireless network designs capable of real-time optimization under adversarial conditions.
     \begin{itemize}[topsep=0pt]
        \item We define and introduce the novel concept of ``Shadow Wireless Intelligence''. Unlike traditional methods that rely on predefined mathematical models to characterize the elements of CC and struggle to cope with dynamic adversarial conditions and environmental changes, SWI can automatically identify detection risks and channel information.  It dynamically adjusts the optimization model in response to evolving adversary strategies and optimizes the balance between covertness and transmission efficiency in real time, effectively addressing the challenges of complex real-world environments.

        \item We propose and detail a structured methodology that efficiently combines DeepSeek-R1’s reasoning capability with real-time, domain-specific knowledge retrieval to dynamically optimize CC strategies. This methodology leverages the strengths of both components to achieve superior performance in complex scenarios.
        % 原文注释掉了, 批注说这个版本所提到的在其他无线优化中也能说
        %\item Finally, we validate the effectiveness of the proposed framework through extensive comparative experiments focusing on optimal AN power control in a full-duplex CC scenario. 
        \item Finally, we validate the effectiveness of the SWI framework through extensive comparative experiments focusing on a full-duplex CC scenario. Specifically, we use SWI to derive the optimal AN power to achieve maximum covert rate under the given covertness constraint. Through benchmarking against leading models (e.g., GPT-o1-mini, Gemini 2.0), we demonstrate that DeepSeek-R1 achieves superior performance in covertness-rate tradeoff.
    \end{itemize}
    
    \section{OVERVIEW OF LLM-DRIVEN REASONING IN CC}
SWI is a novel architectural paradigm in secure edge intelligence, one that integrates advanced LLM reasoning, real-time optimization, and semantic-level security with CC. Departing from conventional rule-based or static learning-based approaches, SWI enables adaptive, context-aware decision-making in adversarial environments by synergizing structured reasoning, RAG, and domain-specific optimization. For example, in CC scenarios, where the primary rule is to reduce the likelihood of detection by a passive adversary or ``warden'', the system inevitably faces high-dimensional, non-convex optimization problems, such as maximizing covert rate under stringent power and covertness constraints.

    \subsection{Covert Communications}
In \Cref{fig:covet_communication_overview}, we outline the core concepts of CC, including its practical applications, core challenges, and enabling techniques. CC can be employed in IoT, satellite, and mobile wireless networks. These systems involve a transmitter, a receiver, and an adversary (or warden), intending to hide transmission activities from being detected. Common techniques, such as AN generation, channel manipulation, and intelligent reflecting surfaces, are used to obscure the transmission. However, enhancing covertness often compromises transmission rate. Therefore, a key challenge in CC is to optimize the trade-off between security and transmission efficiency.
    % \begin{figure*}
    %     \centering
    %     \includegraphics[scale=0.5]{img/RAG_key_features.png}
    %     \caption{ RAG provides real-time updated expertise to LLMs and mitigates
    %     the hallucination issues to some extent.}
    %     \label{fig:RAG_features}
    % \end{figure*}
 \begin{figure*}
    \centering
    \includegraphics[width=0.8\linewidth]{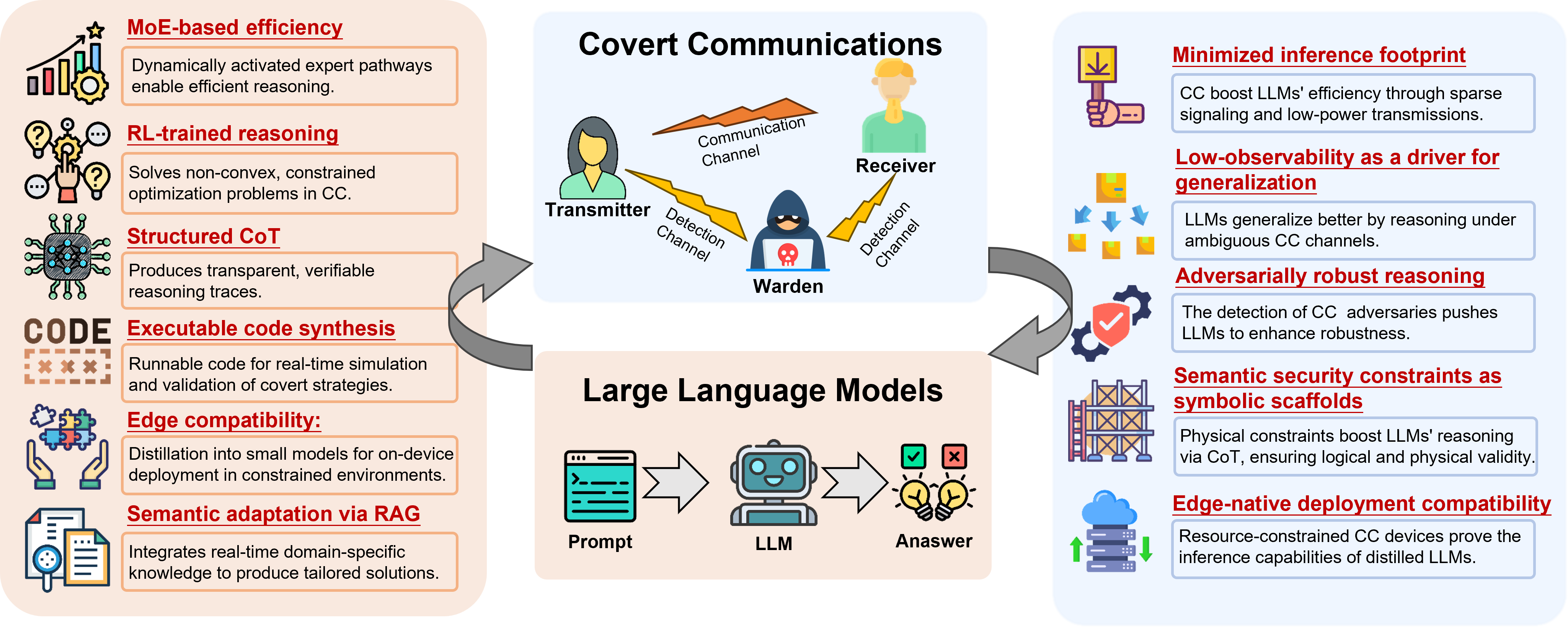}
    \caption{Advantages and benefits: LLM in CC-based optimization and CC for LLM inference.}
    \label{fig:enter-label}
\end{figure*}

\subsubsection{\textbf{LLMs for CC}}
LLMs exhibit powerful reasoning capabilities and knowledge generalization, making them promising tools for addressing the high-dimensional, real-time optimization challenges inherent in CC. In particular, CC systems, designed to hide the very existence of wireless transmission, rely on the precise manipulation of AN, power control, and channel characteristics. However, traditional approaches often fall short in dynamic or adversarial environments due to their reliance on rigid mathematical formulations.
SWI does not attempt to redesign the internal reasoning primitives of LLMs like DeepSeek-R1, which already incorporate RL, MoE, and CoT prompting. Instead, SWI introduces a domain-specific orchestration layer that connects these reasoning capabilities to the physical-layer objectives of CC systems. Specifically, SWI injects real-time measurements (e.g., SNRs or detection error bounds) into the LLM via retrieval-augmented prompts, and validates its symbolic derivations through structured output verification.

This approach allows LLMs to generate optimized CC strategies in real time, including AN power allocation and covert rate tuning, while maintaining adherence to semantic security constraints.
% As shown in our case study, DeepSeek-R1 demonstrates strong performance in deriving closed-form expressions and generating simulation code that balances covertness with throughput.
The benefits of  integrating LLMs with SWI in CC scenarios are summarized below:

\begin{itemize}
  \item \textbf{MoE-based efficiency:} Dynamically selects sparse expert pathways based on input context, reducing reasoning overhead under low observability.
  \item \textbf{RL-trained adaptability:} Solves non-convex, constrained optimization tasks, e.g., covert rate maximization, by leveraging reward-based reasoning.
  \item \textbf{Structured CoT:} Produces stepwise, interpretable symbolic derivations grounded in physical-layer constraints.
  \item \textbf{Executable code synthesis:} Translates optimization logic into runnable MATLAB or Python code for verification and simulation.
  \item \textbf{Semantic alignment via RAG:} Incorporates real-time domain knowledge, e.g., channel models, adversarial thresholds, to improve the relevance of reasoning.
\end{itemize}
%Using in-context knowledge and external retrieval via RAG, they can integrate constraints and parameters into formal models and derive efficient solutions.
\subsubsection{\textbf{CC for LLMs}}

Beyond its traditional role in protecting wireless transmissions, CC provides a unique, high-stakes environment for enhancing LLM inference. In CC systems, transmitters must hide communication activities from adversarial detectors while still achieving functional rate. These stringent requirements introduce sparse, uncertain, and adversarial operating conditions/factors that challenge LLM reasoning and reveal the limits of static generalization.

Key advantages of CC for enhancing LLM performance within SWI include:
\begin{itemize}
  \item \textbf{Inference footprint minimization:} CC’s reliance on sparse signaling aligns with MoE-based routing, reducing token consumption and computational cost.
  \item \textbf{Robustness via adversarial feedback:} The warden’s detection attempts act as real-time adversarial signals, prompting fallback strategies and risk-aware reasoning.
  \item \textbf{Constraint-guided reasoning:} Physical-layer limits such  as the covert rate requirement $R_c \geq \tau$ and the detection error rate constraint $\mathcal{P}_e \geq 1-\xi$ serve as symbolic scaffolds for generating physically valid outputs.
  \item \textbf{Generalization through low observability:} Incomplete or ambiguous channel knowledge forces LLMs to reason beyond memorized distributions, improving out-of-distribution robustness.
  \item \textbf{Edge deployability:} CC systems demand lightweight reasoning under strict latency and energy budgets, fulfilled by distilled LLM variants paired with SWI’s prompt compression and output validation.
\end{itemize}

    \subsubsection{\textbf{SWI Reasoning in CC}}
\begin{figure}[t]
        \centering
\includegraphics[width=\linewidth]{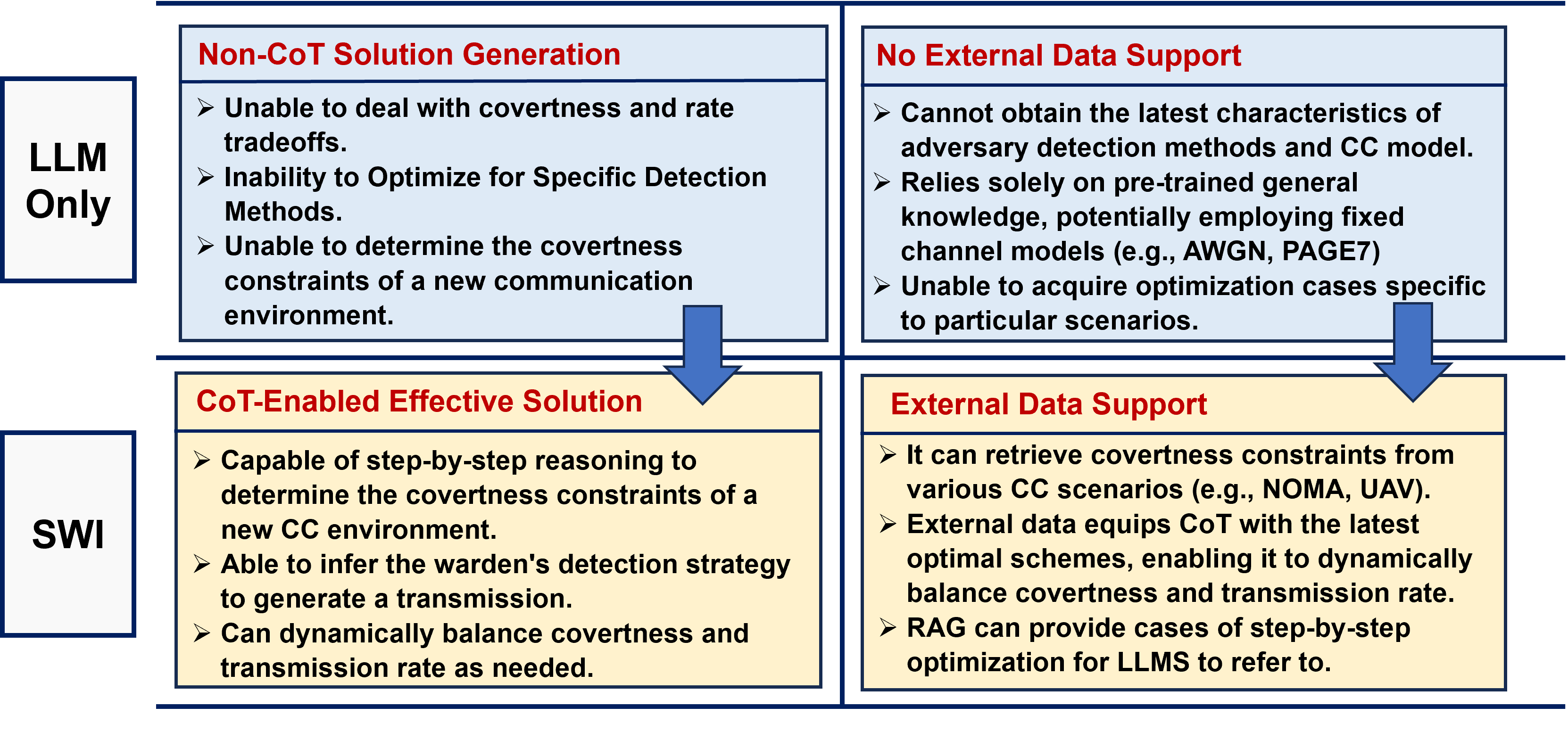}
        \caption{SWI outperforms LLM-only by combining CoT reasoning and external data.}
\label{fig:comparison_method}
    \end{figure}

    As shown in \Cref{fig:comparison_method}, the SWI system enhances the complex reasoning capabilities of LLMs and improves the transparency of the reasoning process by integrating CoT and RAG technologies. This approach also addresses key challenges faced by traditional LLMs: 
    \begin{itemize}
        \item \textbf{Adaptability to new problems}: CoT prompting significantly enhances the adaptability of LLMs in multiple complex reasoning tasks, including symbolic, commonsense, and mathematical reasoning \cite{CoT}. 
        \item  \textbf{Transparent multi-step solutions for complex problems}: By integrating real-time RAG with CoT, the system dynamically retrieves domain-specific knowledge to enhance the reasoning process. This combination not only boosts the expertise of LLMs in handling complex tasks but also improves the accuracy and transparency of their reasoning steps \cite{xu2024cached}.
    \end{itemize}
    \subsection{Shadow Wireless Intelligence Framework}
    The SWI framework includes three modules, i.e., a CC-specific knowledge base, an input/output module and an optimization reasoning solver. 
    
    % \Cref{fig:RAG_enabled_LLM_to_solve_optimal_problem} illustrates the process of solving the CC optimization problem using the SWI framework.

    \begin{itemize}
        \item \textbf{CC-specific nowledge base:}
The knowledge base module is responsible for storing, managing, and retrieving external knowledge.  It converts CC-related information, such as techniques, protocols, historical cases, and optimization strategies, into a vector database for storage and periodically updates the data.  During the reasoning phase, the knowledge base module uses semantic retrieval to provide the LLM with the necessary background knowledge, helping it generate accurate intermediate reasoning steps and final answers.
        \item \textbf{Input/output module:} The input/output module handles use prompts and standardizes the output from the LLM. When a user inputs a prompt, the input processing component matches the optimization problem with relevant content from
        the knowledge base, retrieving reasoning steps and decision variables. The input module then enriches the prompt with domain-specific knowledge to create a CoT-enhanced prompt, which is fed into the LLM for solving.
        During the output phase, the output component uses regular expressions to validate the output format and iterates with the LLM, until the output meets the specified standards.
        \item \textbf{Optimization reasoning solver:} After receiving the enhanced CoT prompt, the solver first conducts a semantic analysis to clarify the optimization objective and identify all constraints.  It then systematically breaks down the overall optimization problem into several interrelated sub-problems based on the problem's structure and logical relationships, ensuring each can be solved independently while maintaining consistency with the overall problem.
For each sub-problem, the solver conducts rigorous validation of the solution's correctness.  Once all sub-problems are successfully solved, their solutions are systematically integrated.  By combining these constraints with the derived results of these sub-problems, the solver ultimately obtain the solution to the entire problem.
    \end{itemize}
    \section{CASE STUDY: LLM-Driven Reasoning in CC}
    In this section, we present the application of the proposed framework to CC-based optimization through a case study. In this study,
    we employ DeepSeek-R1, DeepSeek-V3, Gemini 2.0-Flash, and GPT-o1-mini to determine
    the maximum AN power under a given covert rate. We further demand that the LLMs produce simulation experiment code to illustrate the correlation between the maximum AN power and the transmission outage probability within the CC system. Finally,
    we compare the closed-form expressions generated by these LLMs with the
    solutions of references and evaluate the accuracy of the LLM-generated
    results as well as the effectiveness of the simulation experiment code.
    % \begin{figure*}
    %     \centering
    %     \includegraphics[scale=0.4]{images/CoT_Advantage.png}
    %     \caption{The Optimization Process of LLM with CoT Enhancement.}
    %     \label{fig:RAG_enabled_LLM_to_solve_optimal_problem}
    % \end{figure*}
    % \subsection{Motivation}
    %In covert communication systems, a multitude of optimization challenges exist. Deriving the closed-form expression for optimal power under covert performance constraints demands rigorous mathematical derivation.  Traditional methods usually involve manual formulation and resolution of constraints, such as the establishment of system models and statistical detection models, the establishment of joint optimization objective functions and solutions. This requires a lot of specialized knowledge and advanced mathematical reasoning skills.\\

    \indent
    %With the increasing demand for secure communication in fields such as military, finance, and privacy protection, traditional encryption technologies are struggling to defend against deep traffic analysis attacks because they inherently reveal the act of communication.
    %Covert communication, which hides the existence of signal transmission,
    %offers a new paradigm to counter active detection and interference. 

  CC primarily exploits channel fading or transmission power variations to manipulate the warden's received signals, thereby achieving covert transmission \cite{Covert_Communication_comprehensive_survey}. However, these methods have limitations in ensuring effective covertness, as the inherent randomness of channel characteristics may not be sufficient to fully obscure the signals. To enhance the security of CC, researchers have incorporated AN to further improve covertness. Integrating AN into signal transmission significantly  reduces the warden's ability to distinguish between target signal and background noise, but it also presents new challenges:
   
    \begin{itemize}
        \item \textbf{Noise power control}: Excessive AN increases interference cost and effect legitimate users' reception quality. The AN power must be precisely controlled to confuse detectors without causing excessive interference to legitimate users' communications.
        \item \textbf{Transmission capacity limitation}: 
The transmission capacity of CC follows the square root law, i.e., the amount of information transmitted is proportional to the square root of the number of channel uses to keep a low probability of detection \cite{SRL}. Hence, when spectrum resources are 
limited, the covert transmission efficiency decreases.
        \item \textbf{Covertness-rate trade-off}: High covertness requirements constrain achievable communication rates. Low power transmission of the signal may cause the communication rate to decrease, balancing covert rate and covertness performance is a challenge \cite{tradeoff}.
    \end{itemize} \indent
    \indent

    %Conventional optimal methods struggles to achieve precise regulation in complex channel environments, often causing legitimate user interference or insufficient covertness while failing to balance the covertness-rate tradeoff.  This necessitates innovative methods for reliable covert communication with dynamic tradeoff solution.  Recent advances in inference-oriented LLMs (e.g., DeepSeek-R1, OpenAI o1) demonstrate promising solutions through their language comprehension, mathematical reasoning, and RAG-enhanced knowledge retrieval capabilities.  As illustrated in \Cref{fig:comparison_method}, we list the advantages and disadvantages of different methods in solving the covert communication optimization problem.
  To address these challenges, innovative methods are needed to achieve reliable CC with dynamic trade-off solutions.   Recent advances in inference-oriented LLMs, such as DeepSeek-R1 and OpenAI o1, have demonstrated significant potential due to their capabilities in linguistic understanding and mathematical reasoning.   By synergistically combining the robust reasoning abilities of these LLMs with the CoT and RAG, we can provide an effective approach to tackle these challenges.   This case study leverages these technological advantages by employing DeepSeek-R1 to verify the effectiveness of inference-oriented LLMs in CC-based optimization and simulation code generation.
    % \Cref{fig:RAG_enabled_LLM_to_solve_optimal_problem}, shows the framework used in this Case Study, it includes a domain-specific knowledge base storing optimization literature, a context-aware retrieval augmentation module that converts knowledge into a searchable vector index and retrieves relevant context for user questions, and an intelligent optimization reasoning solver using LLMs to generate solutions. In this framework, our case study assesses the feasibility of LLMs in solving covert communication optimization problems. This assessment is done by comparing the accuracy and code usability of DeepSeek-R1 and o1-mini in covert communication optimization and simulation code generation.
 
    \subsection{Optimal Power Derivation Experiment}
    \subsubsection{\textbf{Problem Description}}
    In the experiment, we consider the same wireless communication setup
    in \cite{full-duplex}, where the receiver (Bob) operates in full-duplex mode,
    while the transmitter (Alice) aims to covertly transmit information to Bob with the aid of AN generated by Bob. Meanwhile, the
    adversary (Willie) attempts to detect this covert transmission. \\
    \indent
    We consider the worst-case scenario where Willie's expected detection error rate reaches its minimum. Our goal is to use SWI to derive the optimal closed-form expression of Bob's maximum power under the constraint of a given covert rate. We also
    tested the availability of LLMs simulation experiment code generation in the
    field of CC.
    \begin{figure*}[htbp]
        \centering
        \includegraphics[width=0.8\textwidth]{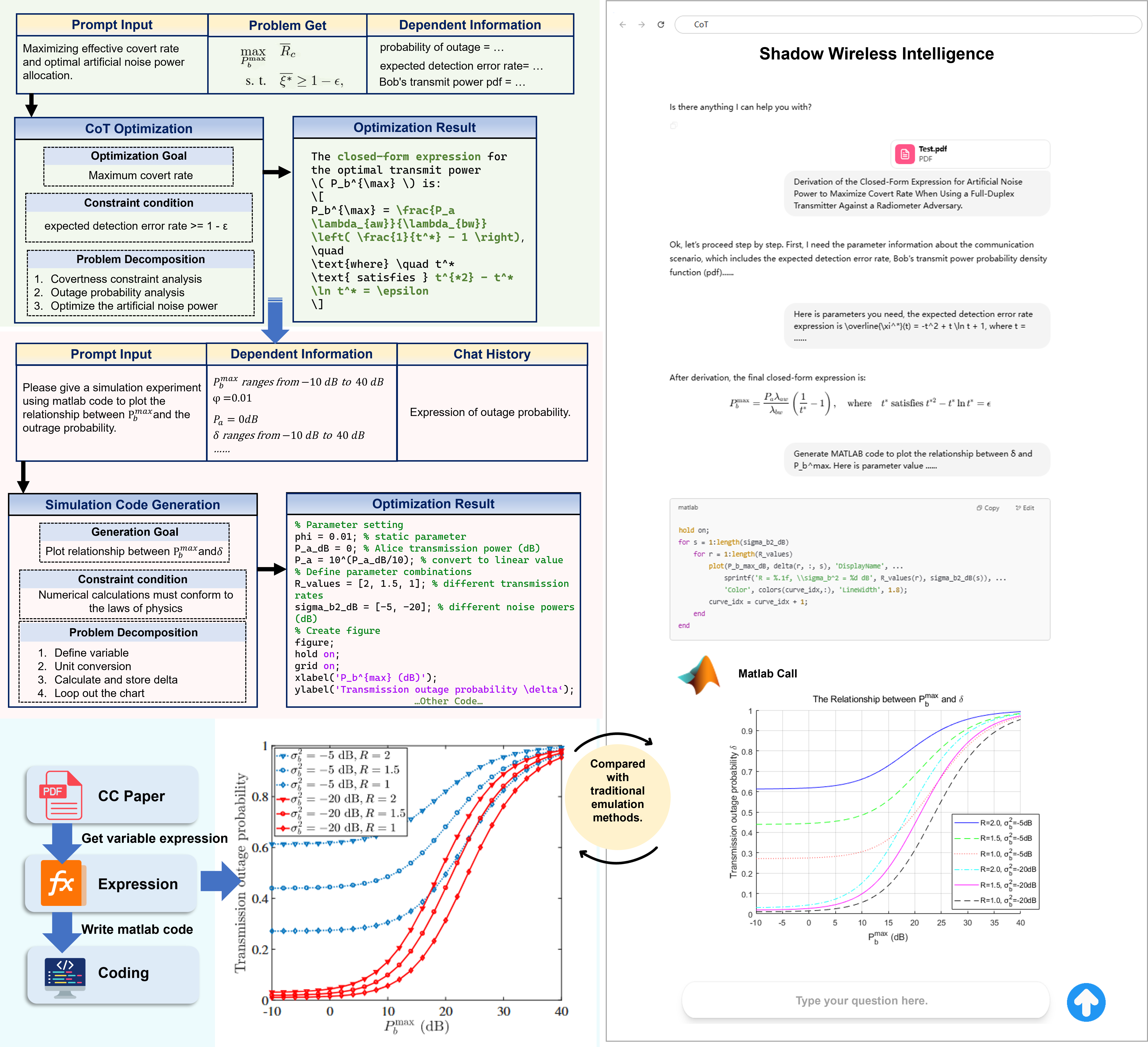}
        \caption{Workflow of SWI for optimal noise power under covertness constraints.}
        \label{fig:workflow}
    \end{figure*}
\subsubsection{\textbf{Experimental Settings}}
    We have downloaded the required references on CC from the IEEE database to serve as a data source for subsequent RAG operations. Based on the reference, we designed Few-Shot-CoT prompts for the LLMs \cite{CoT}. To derive the optimal closed-form expression of Bob's power, we divided the problem into several sub-problems: analyzing transmission outage probability, handling constraints related to the expected detection error rate, examining the distribution characteristics of Bob's power, and solving the optimization problem.  In the process of constructing vector indexes for RAG, we utilized the LangChain framework to segment the source documents and employed OpenAI's text-embedding-ada-002 model to convert the segmented text into a vector knowledge base. Through extensive parameter optimization experiments, we determined the text segmentation parameters to be a chunk size of 500 tokens and an overlap of 300 tokens. This configuration optimally balances information integrity and retrieval efficiency in vector space representation.

\begin{figure*}
        \centering
        \includegraphics[width=0.9\linewidth]{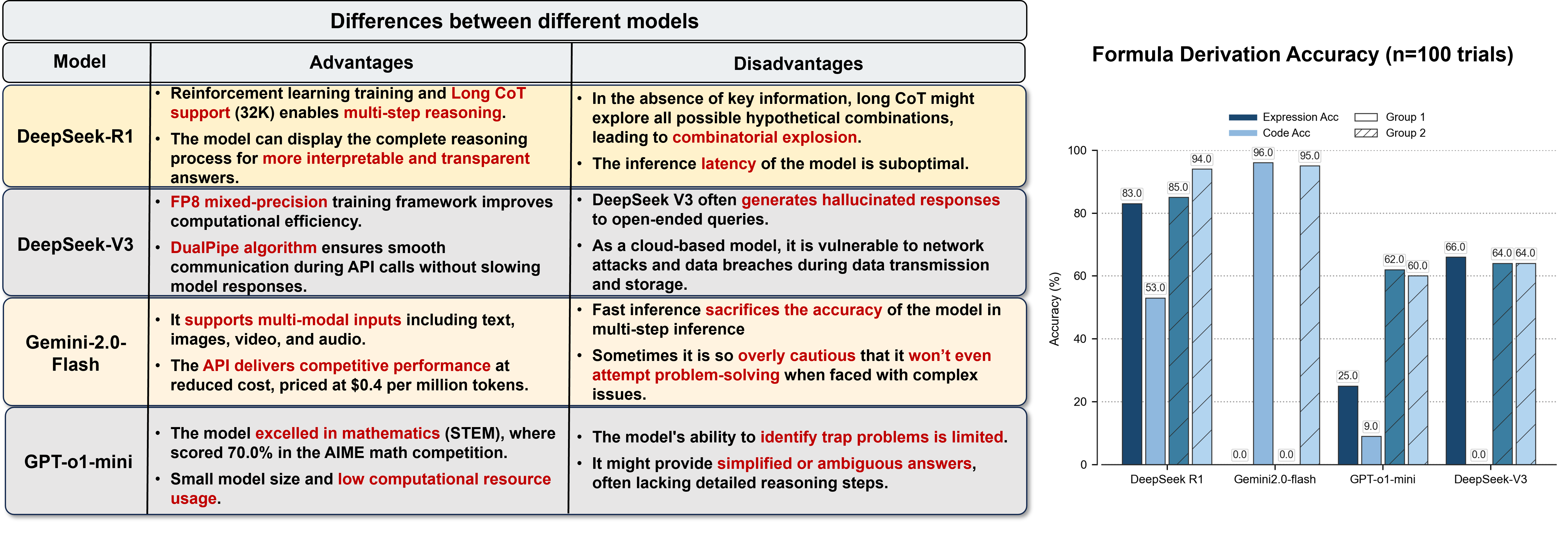}
        \caption{Comparison of Model Advantages, Disadvantages, and Reasoning Precision in the Case Study.}
        \label{fig:diff_between_models}
    \end{figure*}

 \indent
    The derivation of closed-form expressions for Bob's power under specific covertness constraints was conducted using DeepSeek-R1, DeepSeek-v3, Gemini-2.0-Flash and GPT-o1-mini, respectively. \Cref{fig:diff_between_models} compares the advantages and disadvantages of the four models used in the case study. To ensure deterministic outputs and minimize stochastic variations in generated content, the temperature parameter was set to 0 throughout all experimental procedures. \\
    \indent
    To facilitate information extraction, a specific format is designed for the responses generated by LLMs. However, failure to generate content in the required format would prevent the output from being extracted and utilized. To address this, we designed two distinct groups for comparative analysis:

    \begin{itemize}
        \item \textbf{Group 1:} 
The LLM generates content based solely on the prompt, without any response format detection. If the response does not match the specified format, it will be deemed incorrect during performance evaluation.
        \item \textbf{Group 2:} Check the reply format of the LLM. If the LLM reply format does not meet requirements, the formatting detection function will
            instruct the LLM to regenerate the content in compliance with the
            required format.
    \end{itemize}

    \indent
For each set, we instructed the LLM to derive a closed-form expression and provide MATLAB code to simulate the relationship between outage probability and Bob's maximum power.
    \subsubsection{\textbf{Problem-Solving Process}}
    As shown in the \Cref{fig:workflow}, to address solve the above optimization
    problems, we use LLM and CoT processing flow:
    \begin{itemize}
        \item \textbf{Problem analysis:}
       Upon receiving the user’s prompt ``Maximizing the effective covert rate and optimizing the allocation of Bob's power.'', the LLM semantically analyzes it and concludes that the optimization problem is to maximize Alice's effective covert rate while guaranteeing its covertness constraint. Once the problem is determined, the LLM analyzes it and breaks down the optimization problem as follows: determining the decision variables (Bob's maximum power), extracting the involving constraints (covertness constraint), and format requirements for output results.
        % User's prompt: ``Maximizing Effective Covert Rate and Optimal Artificial Noise Power Allocation". The LLM identifies that the objective of the CC-based optimization problem is to maximize the effective covert rate while satisfying the constraints of covertness. 
        % Specifically, it can be broken down as follows:
        % The optimization objective is to maximize the effective covert rate.
        % The constraint is that the expected detection error rate needs to meet a certain threshold.
        % The dependent variables include the transmission outage probability, the expected detection error rate, and Bob's transmission power. In this process, LLM feedback dependent variable SWI system uses RAG to retrieve the variable expression in the paper and feedbacks to LLM.
        \item \textbf{Step-by-step reasoning: }Following the problem analysis, the LLM begins to reason systematically, step by step. First, it analyzes the monotonicity of transmission outage probability and finds that the probability decreases as Bob's maximum power increases. Next, it captures the monotonicity of the expected detection error rate and finds that the error rate also decreases as Bob's maximum power increases. Finally, it combines two relationships (i.e., Alice's outage probability and Willie's detection error rate with the varing maximum power of Bob),  to derive the optimal maximum power.
        \item {\textbf{Derivation results output: }} After deriving the optimized result, the LLM retains both the CoT reasoning process and the final inference outcome, and assigns labels to them. This enables the SWI to automatically extract the optimization results from the LLM and process them. The processed output is then classified by a format classifier. If the output can be correctly extracted and used, the SWI will output the extracted results; otherwise, the LLM will be instructed to regenerate the content based on the specified format.
        % Step 3: The LLM combines the optimization objective and constraints. The goal of the optimization problem is to find Bob's power that meets the constraints, maximizing the effective covert rate.
        % \item \textbf{Deriving the optimal solution: }The LLM analyzes the monotonicity of the transmission outage probability and the expected detection error rate.  It finds that both decrease as Bob's power increases.  Since the effective covert rate increases as the outage probability decreases, the LLM solves for the parameters that meet the constraints.  The optimal power is determined by the covert constraints.
    \end{itemize}

  We optimize the dialogue management of the LLM solver with several key strategies. First, structured templates are adopted to standardize the LLM’s output format, with regular expressions that can accurately extracting important information from the generated content. Second, to preserve important dialogue history, the system keeps the original user queries and distilled core responses, while discarding intermediate reasoning steps. This approach improves efficiency by reducing the context length through data compression and helps reduce LLM hallucinations by focusing on relevant, high-quality outputs.
    \subsection{Experimental Results}

    As shown in \Cref{fig:diff_between_models}, within group 1,  DeepSeek-R1 achieves an accuracy of 83\% in closed-form expression derivation with only 1 formatting error, but its accuracy in simulation code generation drops to 53\% due to 41 formatting errors. Gemini-2.0-Flash fails to derive closed-form expressions but achieves a 96\% accuracy in simulation code generation with zero formatting errors. GPT-o1-mini performs poorly in both tasks, with an accuracy of 25\% in closed-form expression derivation and a correctness of 9\% in simulation code generation, accompanied by high response randomness and frequent formatting errors. DeepSeek-V3 has a significant performance gap in the two tasks: it achieves a 66\% accuracy in closed-form expression derivation, but its simulation code generation is invalid due to severe formatting issues.
    In the second group, DeepSeek-R1 achieves a significant improvement in simulation code generation accuracy to 94\% under strict formatting detection, while maintaining its closed-form expression performance at 85\%. Gemini-2.0-Flash retains its 96\% simulation code accuracy but shows no improvement in closed-form expression derivation. GPT-o1-mini reduces formatting errors, increasing its closed-form expression and simulation code correctness to 62\% and 66\%, respectively, but its high response randomness still affects overall performance. With the assistance of a formatting classifier, DeepSeek-V3 maintains its closed-form expression derivation accuracy (66\%) and improves its simulation code correctness to 64\%. Focusing solely on the cost of API inference, DeepSeek-R1 achieves a total cost of merely \$1.03 in our experiment of solving 100 problems, which is significantly lower than that of o1-mini (\$2.59). This makes DeepSeek-R1 the optimal solution for CC systems that require high analytical precision and reliable execution.
    \section{Conclusions and Future Directions}
    This article has introduced SWI, a novel framework that secures data transmission of CC-based AI applications. It has combined the reasoning capabilities of the MoE-based model DeepSeek-R1 with real-time knowledge retrieval through RAG. Traditional implementation methods of CC like AN injection and power control have had limitations—they have struggled to adapt dynamically in complex environments due to high-dimensional optimization challenges. The proposed system has addressed these limitations by building a domain-specific knowledge base and integrating context-aware retrieval with LLM-driven optimization, enabling intelligent real-time decisions for covert transmission strategies. Tests on full-duplex CC scenarios have shown that DeepSeek-R1 substantially outperforms baseline models, achieving 85\% accuracy in closed-form derivation and 94\% in simulation code generation. These results have demonstrated its potential as a foundation for secure and autonomous wireless systems.

    Future research can explore the application of SWI in multi-user scenarios or more general wireless communication scenarios.
    Emerging techniques, such as graph RAG and distributed LLM reasoning, can be employed to improve the inference speed and achieve the optimal resource allocation. Additionally, addressing the balanced allocation of communication and computational resources in these wireless scenarios, while ensuring personalized service delivery to individual user, is essential.
    \bibliography{references}
    \bibliographystyle{IEEEtran}
    \vspace{-33pt}
\begin{IEEEbiographynophoto}
{Yuanai Xie} (IEEE Member) is currently a lecturer at the College of Computer Science, South-Central Minzu University, Wuhan, China. He received the Ph.D. degree in control
science and engineering in Yanshan University, Qinhuangdao, China. His current research interests include vehicular networks and covert communications.
\end{IEEEbiographynophoto}
   \vspace{-33pt}
\begin{IEEEbiographynophoto}
{Zhaozhi Liu} received the bachelor’s degree from Wuhan Huaxia Institute of Technology, Wuhan, China, in 2024. He is pursuing a Master's degree at the School of Computer Science of South-Central Minzu University. His research interests include covert communication, LLM-empowered wreless communication. \end{IEEEbiographynophoto}
   \vspace{-33pt}
\begin{IEEEbiographynophoto}
{Xiao Zhang} received the Ph.D. degree from Department of Computer Science in City University of Hong Kong, Hong Kong, 2016. Currently, he is associate professor with the College of Computer Science, South-Central Minzu University, China. His research interests include wireless and UAV networking, algorithms design and analysis, and combinatorial optimization. 
\end{IEEEbiographynophoto}
   \vspace{-33pt}
\begin{IEEEbiographynophoto}
{Shihua Zhang} received the PhD degree in bioinformatics from Harbin Medical University in 2010.
 He is currently a Profes-
sor with the College of Computer Science, South-
Central Minzu University, Wuhan. His research interests include
machine learning, computational systems biology, and bioinformatics.
\end{IEEEbiographynophoto}
   \vspace{-33pt}
\begin{IEEEbiographynophoto}
{Rui Hou} received
the Ph.D. degree from the Huazhong University of Science
and Technology, Wuhan, China, in 2006. He is currently a Professor with the College of
Computer Science, South-Central Minzu University, Wuhan. His main research interests include
computer network architectures, optical switching, and wireless sensor networks. 
\end{IEEEbiographynophoto}
\vspace{-33pt}
\begin{IEEEbiographynophoto}{Minrui Xu}
received the B.S. degree from Sun
Yat-sen University, Guangzhou, China, in
2021. He is currently working toward the
Ph.D. degree at the School of Computer Science and Engineering, Nanyang Technological University, Singapore.
His research interests mainly focus on
metaverse, deep reinforcement learning,
and mechanism design.
\end{IEEEbiographynophoto}
   \vspace{-33pt}
\begin{IEEEbiographynophoto}
{Ruichen Zhang} (IEEE Member) is currently working as a Post-Doctoral Research Fellow with the College of Computing and Data Science, Nanyang Technological University, Singapore. He received the Ph.D. degree from Beijing
Jiaotong University, China, in 2023. He
is the Managing Editor of IEEE TNSE. His research interests include LLM-empowered networking, generative AI models, and heterogeneous networks.
\end{IEEEbiographynophoto}
   \vspace{-33pt}
\begin{IEEEbiographynophoto}
{Dusit Niyato} (IEEE Fellow) is a professor in the College of Computing and Data Science, at Nanyang
Technological University, Singapore. He received
B.Eng. from King Mongkuts Institute of Technology
Ladkrabang (KMITL), Thailand and Ph.D. in Electrical and Computer Engineering from the University
of Manitoba, Canada. His research interests are in
the areas of mobile generative AI, edge intelligence,
decentralized machine learning, and incentive mechanism design.
\end{IEEEbiographynophoto}

\end{document}